\documentclass[letters,usenatbib]{mn2e}
\usepackage[letterpaper,margin=0.8in]{geometry}
\usepackage{hyperref}
\usepackage{natbibmnfix,fixltx2e}
\usepackage{astrojournals}
\usepackage{amsmath,amssymb}
\usepackage{txfonts}
\usepackage{graphicx}
\usepackage{microtype}

\hypersetup{%
  pdftitle={Accelerating Magnetized Gas Clouds},
  pdfauthor={\textcopyright\ authors},
  bookmarksopen=true,
  colorlinks=true,
  linkcolor=black,
  citecolor=black,
  urlcolor=black}

\RequirePackage{etoolbox}
\newbool{smallfigone}
\setbool{smallfigone}{false}

\date{}
\pagerange{\pageref{firstpage}--\pageref{lastpage}}
\pubyear{2014}

\renewcommand*{\vec}[1]{\boldsymbol{#1}}

\begin{document}
\label{firstpage}

\title[Accelerating Magnetized Gas Clouds]{Magnetized Gas Clouds can Survive Acceleration by a Hot Wind}
\author[McCourt, O'Leary, Madigan, \& Quataert]
{Michael McCourt\thanks{mkmcc@astro.berkeley.edu},
Ryan O'Leary,
Ann-Marie Madigan\thanks{Einstein Postdoctoral Fellow},
 \&
Eliot Quataert \vspace{0.1in} \\
Astronomy Department and Theoretical Astrophysics Center, University of California, Berkeley, CA 94720, USA
}
\maketitle

\begin{abstract}
We present three-dimensional magnetohydrodynamic simulations of magnetized gas clouds accelerated by hot winds.
We initialize gas clouds with tangled internal magnetic fields and show that this field suppresses the disruption of the cloud: rather than mixing into the hot wind as found in hydrodynamic simulations, cloud fragments end up co-moving and in pressure equilibrium with their surroundings.
We also show that a magnetic field in the hot wind enhances the drag force on the cloud by a factor $\sim (1+v_{\text{A}}^2/v_{\text{wind}}^2)$, where $v_{\text{A}}$ is the Alfven speed in the wind and $v_{\text{wind}}$ measures the relative speed between the cloud and the wind.
We apply this result to gas clouds in several astrophysical contexts, including galaxy clusters, galactic winds, the Galactic center, and the outskirts of the Galactic halo.
Our results can explain the prevalence of cool gas in galactic winds and galactic halos and how such cool gas survives in spite of its interaction with hot wind/halo gas.
We also predict that drag forces can lead to a deviation from Keplerian orbits for the G2 cloud in the galactic center.
\end{abstract}


\section{Introduction}
\label{intro}
Many astrophysical scenarios involve the motion of dense gas clouds through hotter, more tenuous surroundings.
Examples include supernova remnants, filaments of cool gas in galaxy clusters, high-velocity clouds (HVCs) of atomic gas in the Galactic halo, and multiphase galactic winds.
More recently, the discovery of G2, a possible gas cloud on a highly eccentric orbit about the massive black hole Sgr~A$^{*}$, has focused interest on the dynamics of gas clouds in the Galactic center \citep{Gillessen2012}.

Cloud-wind interactions have been studied most extensively in the context of the interstellar medium
\citep[e.\,g][]{Klein1990,Stone1992,Klein1994,MacLow1994,Shin2008,Li2013,Johansson2013}. This scenario typically features a fairly low-density cloud ($\rho_{\text{cloud}}/\rho_{\text{wind}} \sim 10$) overrun by an extremely supersonic shock ($M \gtrsim 10$).
The evolution becomes self-similar in this limit \citep{Klein1994,MacLow1994}, and the cloud disrupts on a ``crushing'' timescale equal to the shock-crossing time:
\begin{align}
  t_{\text{crush}} \sim
  \left(
    \frac{\rho_{\text{cloud}}}{\rho_{\text{wind}}}
  \right)^{1/2}
  \frac{R_{\text{cloud}}}{v_{\text{wind}}},
\end{align}
where $R_{\text{cloud}}$ is the radius of the cloud and $v_{\text{wind}}$ is the relative velocity between the cloud and the shock.
Disruption by the Kelvin-Helmholtz and Rayleigh-Taylor instabilities yield similar timescales.
Cooling may extend the cloud's lifetime by a factor of several; however it does not appear to halt its destruction indefinitely \citep{Cooper2009}.

The role of magnetic fields in cloud disruption has not been thoroughly investigated.
Magnetic forces may accelerate the destruction of gas clouds, at least for certain strengths and configurations \citep{Gregori1999,Gregori2000,Li2013}.
We note, however, that these simulations have only explored clouds threaded by straight magnetic field lines (still susceptible to shear and Rayleigh-Taylor instabilities), or clouds with internal magnetic configurations prone to ``pinch''-type instabilities.
We show here that other arrangements can yield qualitatively different results.

In addition to the supernova literature, a number of recent studies have specifically simulated the motion of the G2 gas cloud through the dilute, X-ray emitting plasma permeating the Galactic center \citep{Burkert2012,Schartmann2012,Ballone2013,Anninos2012,Abarca2014,Saitoh2014,Guillochon2014}.
While these studies explore larger density contrasts ($ \rho_{\text{cloud}}/\rho_{\text{wind}} \gtrsim 50$) and lower speeds ($M \lesssim 2$), they largely ignore the role of magnetic fields.\footnote{\citet{Sadowski2013b} include magnetic fields in their simulations; their study focuses on the properties of the bow shock ahead of G2, however, with insufficient resolution to capture the detailed disruption of the cloud.} Neglecting magnetic fields is typically not a well-motivated approximation, however:  in many astrophysical environments, magnetic stresses strongly modify the instabilities that break up clouds in hydrodynamic simulations.

\ifbool{smallfigone}{\begin{figure}}{\begin{figure*}}
\ifbool{smallfigone}{\includegraphics[width=\columnwidth]{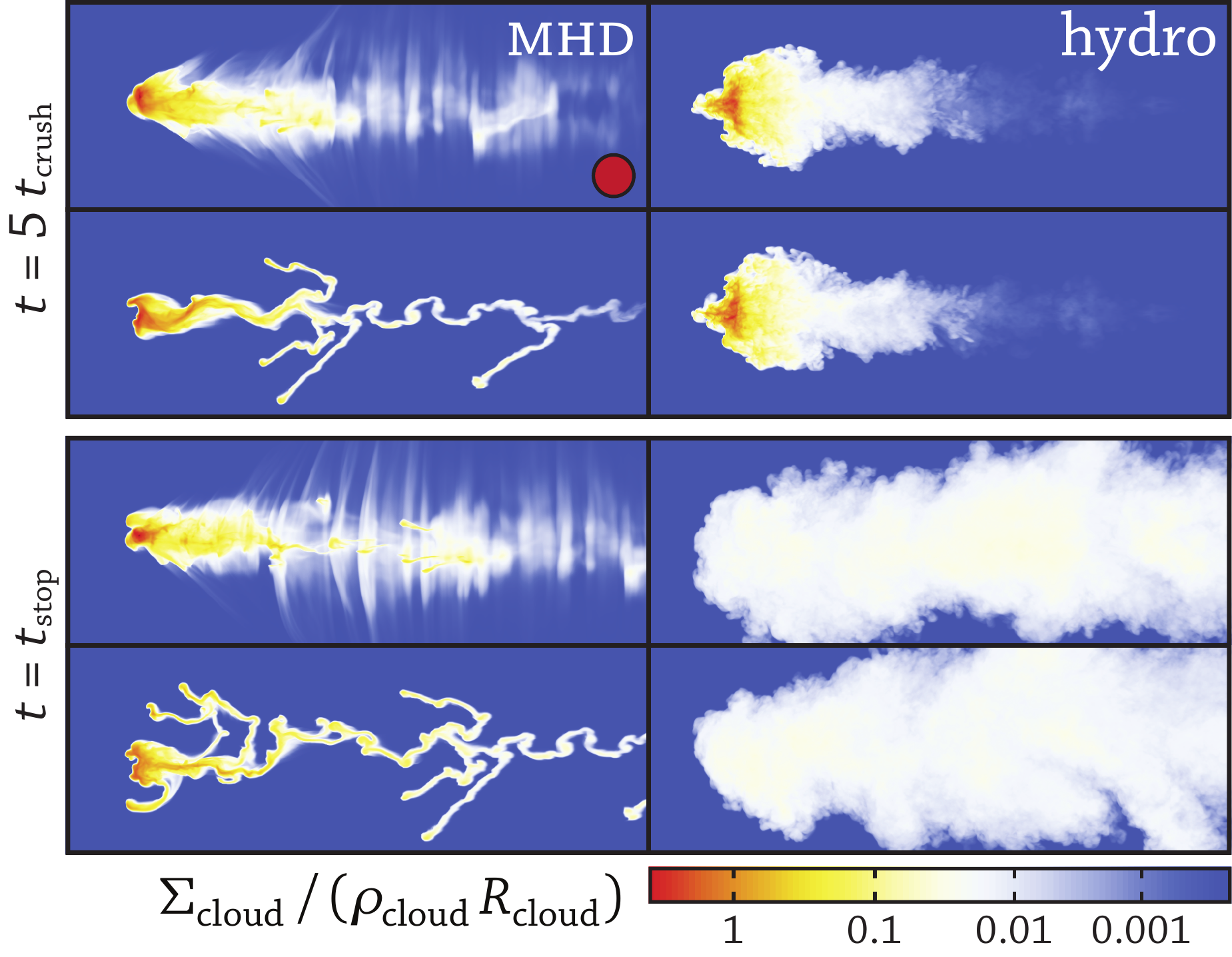}}{\includegraphics[width=\textwidth]{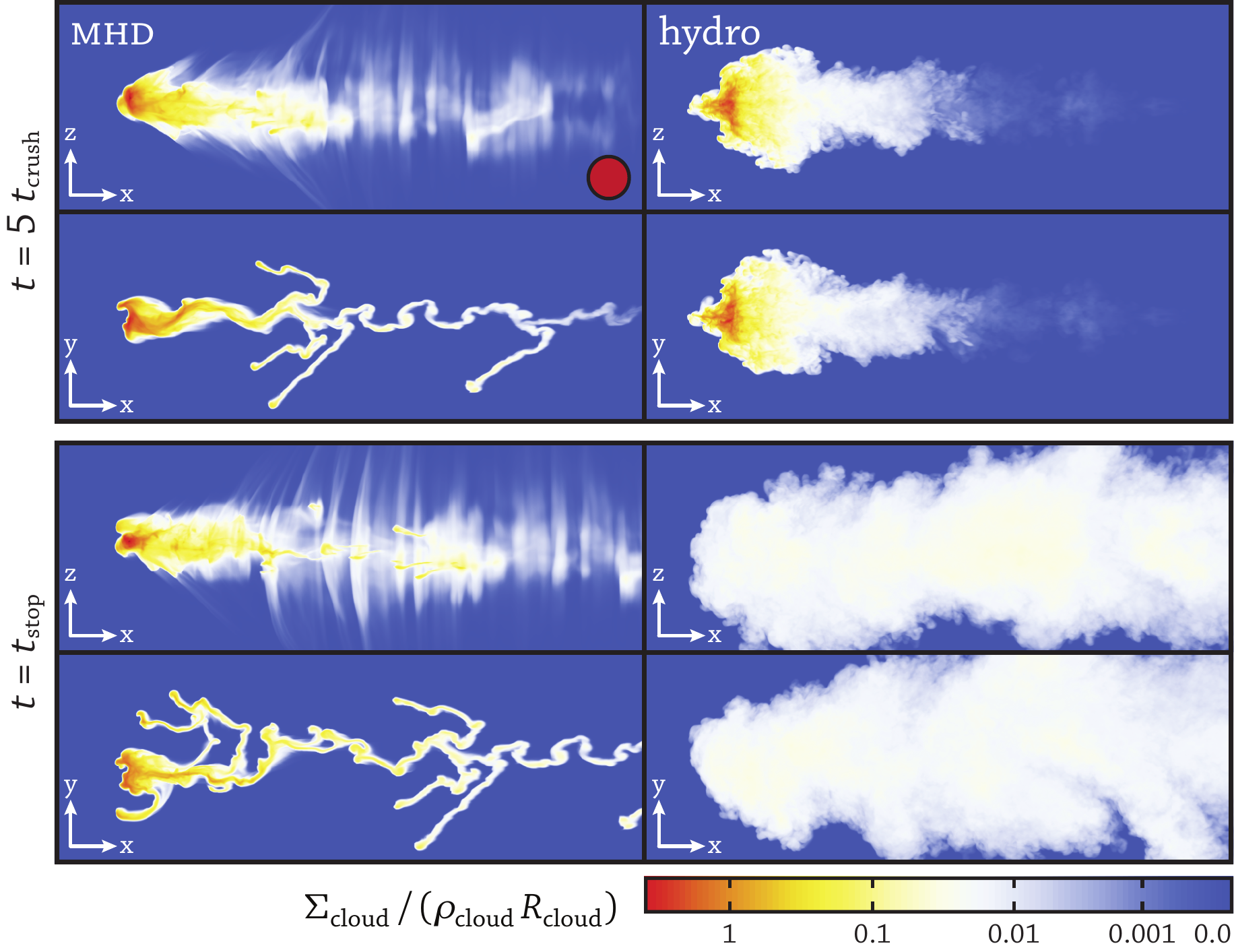}}
  \caption{Comparison of cloud disruption in hydrodynamic and MHD simulations.
Color shows the projected cloud density in our simulations at the time $t = 5\,t_{\text{crush}}$ (\textit{top}) and $t = t_{\text{stop}}$ (\textit{bottom}); the latter is defined as the time when the cloud's velocity relative to the wind drops to about 10\% of its initial value (see figure~\ref{fig:stopping-distance}).
The color scale is logarithmic, and is held constant across all images.
The red circle in the top-left panel shows the initial size of the cloud.
In each of the two groups of images, the upper panel shows the projection along the $y$-axis, orthogonal to the mean magnetic field; the lower panel shows the projection along $z$-axis, down magnetic field lines.
In the hydrodynamic case (\textit{right}), the cloud disrupts entirely and mixes into its surroundings; the results are also statistically isotropic and independent of viewing angle.
In our MHD simulations (\textit{left}), magnetic fields strongly inhibit the disruption; while the cloud breaks into several smaller clumps, these ``cloudlets'' do not break up further or mix into the surroundings.
We also find the MHD results are strongly anisotropic with respect to the magnetic field of the wind (c.f. top and bottom in each group).}
 \label{fig:mhd-hydro-comp}
\ifbool{smallfigone}{\end{figure}}{\end{figure*}}

In this letter, we study the dynamics of cloud-wind interactions using three-dimensional magnetohydrodynamic (MHD) simulations, with higher density contrasts and slower velocities than typically explored in the ``cloud-crushing'' literature.
This parameter regime is relevant for emission-line filaments in galaxy clusters, multiphase galactic winds, HVCs, and gas clouds in the Galactic center.
We study how tangled magnetic fields internal to the cloud influence its evolution, and we show that certain magnetic configurations can in fact suppress the disruption of gas clouds \citep[see also][]{Ruszkowski2007}.
We furthermore show that the drag force experienced by the cloud is sensitive to the magnetic field in the external medium and can in some cases greatly exceed the hydrodynamic estimate \citep[see also][]{Dursi2008}.
Our results generally show that magnetized gas clouds can live longer, and are more strongly coupled to their environments, than hydrodynamic simulations imply.
This letter presents two of our primary results.
A more thorough exploration of parameters and magnetic field configurations will be presented in a forthcoming paper (O'Leary et al., in prep).

\section{Method}
\label{sec:method}
We integrate the MHD equations using the conservative code Athena \citep{Stone2008, Gardiner2008}.
Following \citet{Shin2008}, we use a passively advected scalar to keep track of the cloud and we boost the simulation domain after every time-step to keep the cloud from leaving the boundaries.
This technique significantly reduces the computational cost of our simulations for two reasons: first, keeping the cloud near the center of computational grid permits a smaller domain size.
Additionally, implementing cloud-following reduces the truncation errors by minimizing the relative velocity between the cloud and the computational grid; we thus find faster numerical convergence with this technique.

We run our simulations on a cartesian domain with dimensions $(40\times20\times20) \times R_{\text{cloud}}$, where $R_{\text{cloud}}$ is the initial size of the gas cloud.
We run our simulations with 32~cells per cloud radius; this resolves the ``draping layer'' of swept-up magnetic field lines in all of our simulations \citep{Dursi2008}.

We impose a steady wind with $v_x = 1.5\,c_{\text{s}}$ at the upstream boundary, where $c_{\text{s}}$ is the sound speed in the external medium.
This velocity is appropriate for dense clouds free-falling through hot virialized gas (e.g., G2 and HVCs) as well as clouds driven out by galactic winds.
We use an outflow (zero-gradient) boundary condition downstream, and periodic boundary conditions in the directions orthogonal to the flow.
We choose periodic boundary conditions for their simplicity and numerical stability; we place them far enough from the cloud  to not affect its disruption.

In our MHD simulations, we add a constant magnetic field to the wind with a strength defined via $\beta_{\text{wind}} \equiv 8\pi{}P_{\text{wind}}/B_{\text{wind}}^2$ and a tangled magnetic field to the cloud with a strength measured by $\beta_{\text{cloud}}$.
We generate a tangled, approximately force-free magnetic field inside the cloud via a superposition of ten modes which individually satisfy:
\begin{align}
  \vec{B} = \cos(\alpha\,a) \; \hat{\vec{c}}
          + \sin(\alpha\,a) \; \hat{\vec{b}}\label{eq:force-free-term}
\end{align}
where $\hat{\vec{a}}$, $\hat{\vec{b}}$, and $\hat{\vec{c}}$ form a right-handed coordinate system randomly oriented with respect to the computational domain, and $\alpha = 10 /R_{\text{cloud}}$ sets the correlation length of the magnetic field.
We add straight magnetic fields to the wind, aligned with the $z$-axis in our domain.
We run simulations with three different field strengths in the wind: $\beta_{\text{wind}} = $ 0.1, 1, and 10.
We arrange the initial condition so that the cloud is magnetically isolated from the wind; i.\,e. no field lines enter or leave the cloud initially.
To do this, we truncate the magnetic field defined by equation~\ref{eq:force-free-term} at the edge of the cloud while preserving $\nabla\cdot\vec{B} = 0$; this procedure unfortunately breaks the force-free nature of equation~\ref{eq:force-free-term} and prevents us from initializing our simulations in an exact equilibrium.
Our gas clouds consequently tend to unravel from their initial condition; this artificially reduces any stabilizing effect of the magnetic field.

Previous studies have shown that cooling significantly changes the disruption of gas clouds \citep{Miniati1999,Cooper2009}.
We therefore include cooling in our calculation, implemented as a source term in the energy equation.
We adopt an artificial cooling function which simply keeps the cooling time short compared to the ``crushing'' time inside the cloud, and makes the cooling time longer than the flow time in the wind.
The companion paper shows this cooling curve explicitly, and compares it against simulations with more realistic cooling functions.

We initialize the cloud with a smoothed top-hat density profile and in pressure equilibrium with the surrounding gas.
We restrict ourselves to a cloud with density $\rho_{\text{cloud}} = 50\,\rho_{\text{wind}}$; we explore larger density contrasts in the companion paper.

\section{Results}
\label{sec:results}
\subsection{Internal magnetic fields can inhibit disruption}
\label{subsec:breakup}
Figure~\ref{fig:mhd-hydro-comp} compares a hydrodynamic simulation of cloud disruption against an MHD calculation with $\beta_{\text{wind}} = \beta_{\text{cloud}} = 1$.
In the hydro simulation (\textit{right}), a combination of shear and Rayleigh-Taylor instabilities shred the cloud and mix it completely into the surroundings; this is in agreement with previous studies.
The cloud in our MHD simulation does not dissociate so completely, however.
While the initial gas cloud does break up into several smaller ``cloudlets,'' magnetic stresses prevent these smaller clumps from mixing into the surrounding gas before the cloud decelerates and co-moves with the hot wind (see section~\ref{subsec:drag} below).
The final state is a new quasi-equilibrium in which the relative motions between cloud fragments and wind have ceased.
Thus, we find that a tangled magnetic field threading the cloud may forestall its destruction, at least for some configurations.

The resulting density distribution for the MHD simulation is substantially more clumped and less mixed than the gas in the hydrodynamic simulation.
Observable properties of the gas cloud (e.\,g., emission and absorption lines) thus differ significantly between the two cases; hydro simulations therefore cannot be used to predict observable properties of gas clouds with dynamically significant magnetic fields.

The observable properties of the clouds in our MHD simulations also depend on the viewing angle with respect to the large-scale magnetic field.
The top and bottom rows in figure~\ref{fig:mhd-hydro-comp} compare the density distributions projected along (\textit{top}) and across (\textit{bottom}) the large-scale magnetic field in the wind.
The cloudlets and the interstitial gas between them closely trace magnetic field lines; the two projections in figure~\ref{fig:mhd-hydro-comp} thus appear very different.
As expected, the hydrodynamic simulation exhibits no such anisotropy.

We quantify the differences between our hydro and MHD simulations in figure~\ref{fig:clumping-factor}, which shows the ``clumping factor'' $c \equiv \langle \rho_{\text{cloud}}^2 \rangle^{1/2} / \langle \rho_{\text{cloud}} \rangle$ as a function of time in our simulations.
The hydro simulation (\textit{dotted black curve}) mixes into its surroundings on a timescale of order $\sim 10\,t_{\text{crush}}$; this is consistent with previous studies that include radiative cooling \citep[e.\,g.][]{Cooper2009}.
The MHD simulation with $\beta_{\text{cloud}} = 1$ (\textit{solid blue curve}) evolves very differently: after an initial phase in which the cloud readjusts due to the initial conditions and its impact with the surrounding wind, the cloud  enters a long-lived phase in which it does not mix into the surrounding gas (cf. figure~\ref{fig:mhd-hydro-comp}).
We demonstrate the robustness of this result by also including a cloud with a weaker internal field $\beta_{\text{cloud}} = 10$ in figure~\ref{fig:clumping-factor} (\textit{dashed red curve}).
The cloud with $\beta_{\text{cloud}} = 10$ behaves similarly to the one with a stronger internal field, especially at late times.

In our MHD simulations, we find that the cloud breaks up into a small number of clumps; these ``cloudlets'' appear stable and do not break up or mix into their surroundings.
The process determining the size of these stable clumps is essential to understanding our result.
The cloudlet size does not seem to be determined by the numerical resolution, but it may depend on the strength and spectrum of the tangled field initially permeating the cloud.
We explore this in more detail in the companion paper.

\subsection{Enhanced drag force from the external magnetic field}
\label{subsec:drag}
\begin{figure*}
\centering
\begin{minipage}[t]{\columnwidth}
  \centering
  \includegraphics[width=\textwidth]{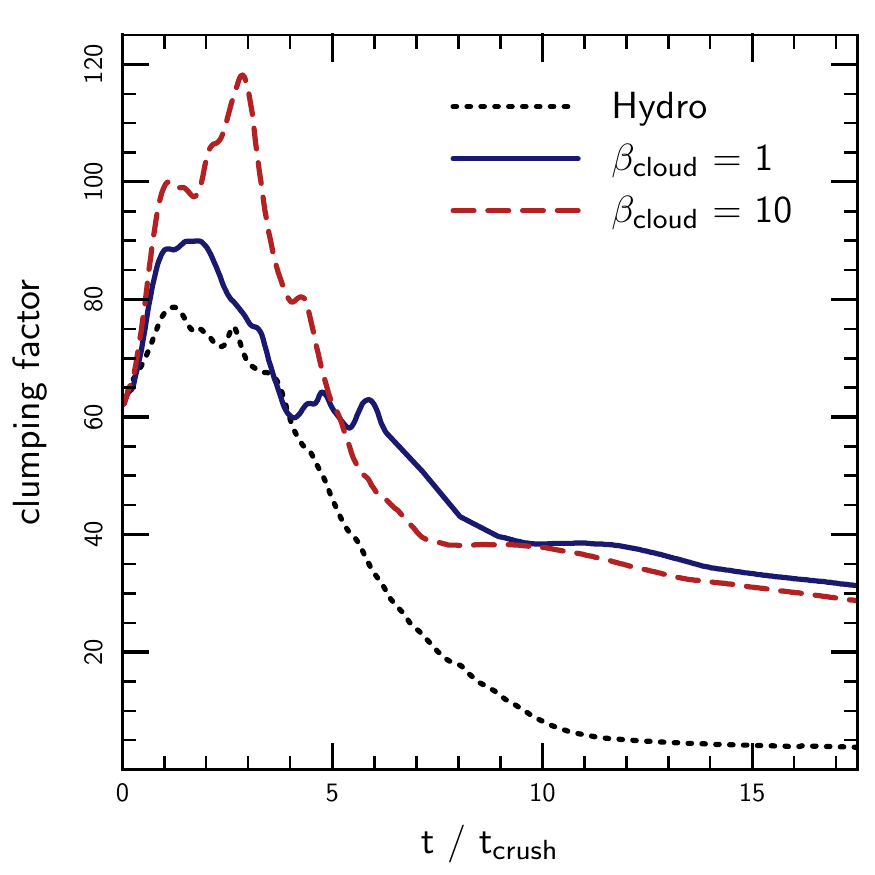}
  \caption{Disruption of gas clouds with different \textit{internal} magnetic field strengths.
We quantify the disruption via the ``clumping factor'' $c \equiv \langle \rho_{\text{cloud}}^2 \rangle^{1/2} / \langle \rho_{\text{cloud}} \rangle$, where $\langle\cdots\rangle$ represents a spatial average over the entire computational domain.
Our initial condition has a clumping factor $c = (V_{\text{domain}}/V_{\text{cloud}})^{1/2} \sim 62$; a clumping factor $c = 1$ indicates a cloud that evenly fills the domain and is fully mixed into its surroundings.
The hydro simulation (\textit{dotted black curve}) mixes into its surroundings on a timescale $\sim\,10\,t_{\text{crush}}$; this is consistent with previous studies.
The MHD simulations evolve very differently (\textit{solid blue curve} and \textit{dashed red curve}): after an initial rearrangement lasting $\sim 5\,t_{\text{crush}}$,  they enter a long-lived phase in which the cloud does not mix into the surrounding gas (cf. figure~\ref{fig:mhd-hydro-comp}).}
  \label{fig:clumping-factor}
\end{minipage}
  \hfill
\begin{minipage}[t]{\columnwidth}
  \centering
  \includegraphics[width=\textwidth]{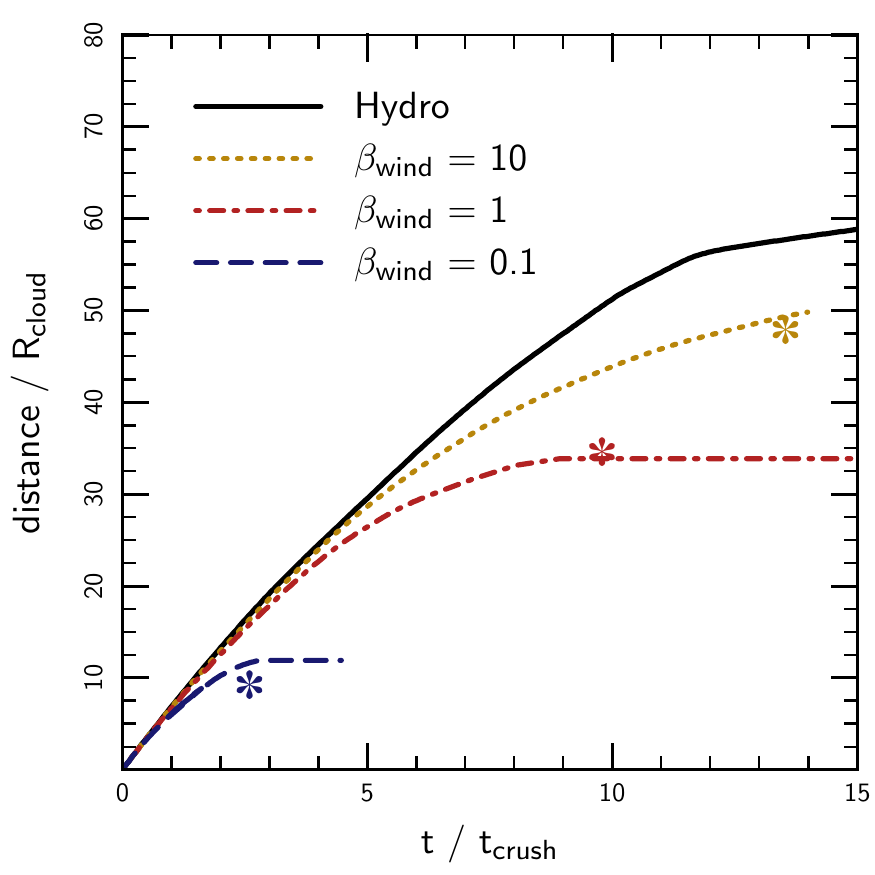}
  \caption{Comparison of the drag force for different {\em wind} magnetic field strengths.
Lines show the total distance traveled by the center of mass of the clouds in a frame co-moving with the wind.
We plot this distance as a function of time for a hydrodynamic simulation, followed by MHD simulations with $\beta_{\text{cloud}} = $ 1 and $\beta_{\text{wind}} = $ 10, 1, and 0.1.
We find that magnetic field with $\beta_{\text{wind}} \sim 1$ roughly doubles the drag force on the cloud.
A much stronger field with $\beta_{\text{wind}} \sim 0.1$ results in a drag force significantly larger than the hydrodynamic drag force.
Asterisks mark `stopping' distances and times estimated using equation~\ref{eq:drag-model}; these are the points at which the cloud should become co-moving with the wind.
These estimates agree well with our simulation data, particularly in the strongly-magnetized limit with $\beta_{\text{wind}} \lesssim 1$.
Lines end either when 5\% of the cloud mass has left the domain or, in the case of $\beta_{\text{wind}} = 0.1$, at $t = 5$ ($\sim 4.5\,t_{\text{crush}}$), when we stopped that simulation.}\label{fig:stopping-distance}
\end{minipage}
\end{figure*}
The drag force is an important process influencing the observable properties of gas clouds: it sets the terminal velocity of the filaments in galaxy clusters and determines whether atomic and molecular clumps in galactic winds co-move with the rest of the outflow.
Moreover, the deceleration of G2 due to its drag force may be directly observable as a deviation from its Keplerian orbit; this could in turn place powerful constraints on the (unknown) gas density and magnetic field strength at radii of 100s of AU in the Galactic center.

When the wind is strongly magnetized ($\beta_{\text{wind}} \lesssim 1$), we find that the drag force exceeds the hydrodynamic estimate, $F_{\text{hydro}} \sim \rho_{\text{wind}} v_{\text{wind}}^2 R_{\text{cloud}}^2$.
Figure~\ref{fig:stopping-distance} illustrates this drag force by showing the distance traveled by the clouds in our simulations: in all cases where $\beta_{\text{wind}} \lesssim 1$, the cloud accelerates much faster than it does in the hydro simulation.

We can understand this enhanced drag force with a simple model.
As the cloud moves through the hot gas, it sweeps up magnetic field lines at its leading edge.
If the magnetic field is initially coherent on a lengthscale larger than the cloud size, the shape of the swept-up field lines is set by a balance between ram pressure and magnetic tension.
The field lines in front the cloud thus bend with a characteristic radius of curvature $R_{\text{curv}} \sim (v_{\text{A}}/v_{\text{wind}})^2 R_{\text{cloud}}$, where $v_{\text{A}} \equiv B/\sqrt{4\pi\rho}$ is the Alfv\'en speed in the wind; the cloud must then accelerate a column of gas with the cross-section $\sim R_{\text{cloud}} \times R_{\text{curv}}$.\footnote{We note that this analysis breaks down when the drag force is extremely large; i.\,e. when $R_{\text{curv}} \gg$ the coherence length of the field.}
The total drag force decelerating the cloud is then given by \citep[cf.][]{Dursi2008}:
\begin{align}
  F_{\text{drag}} \sim \rho_{\text{wind}} v_{\text{wind}}^2 R_{\text{cloud}}^2 \times
\left(1 + \frac{v_{\text{A}}^2}{v_{\text{wind}}^2}\right).\label{eq:drag-model}
\end{align}
The MHD drag force is larger than the hydrodynamic force by a factor of $[1 +(v_{\text{A}}/v_{\text{wind}})^2]$.

The drag force in equation~\ref{eq:drag-model} can also be expressed as a ``stopping distance'' $d_{\text{stop}} \sim M v^2 / F_{\text{drag}}$: this is approximately the distance a cloud moves relative to its surroundings before becoming co-moving.
We find
\begin{align}
  d_{\text{stop}} \sim R_{\text{cloud}}
  \times \left(\frac{\rho_{\text{cloud}}}{\rho_{\text{wind}}}\right)
  \times \frac{\beta_{\text{wind}} M^2}{1 + \beta_{\text{wind}} M^2},\label{eq:dstop}
\end{align}
where $\beta_{\text{wind}} \equiv 8\pi{}P_{\text{wind}}/B_{\text{wind}}^2$ measures the magnetic field strength in the wind and $M \equiv v_{\text{wind}} / c_{\text{s, wind}}$ is the Mach number of the cloud relative to its surroundings.
The first two terms in equation~\ref{eq:dstop} intuitively express that the cloud should stop after intercepting a comparable mass of wind material; this is the usual result from hydrodynamics.
Magnetic stresses contribute the final term in equation~\ref{eq:dstop}; this enhancement is significant when the product $\beta M^2 \ll 1$.
Equation~\ref{eq:dstop} implies stopping distances of 50, 30, and 10~$R_{\text{cloud}}$ for the simulations in figure~\ref{fig:stopping-distance} with $\beta_{\text{wind}} =$ 10, 1, and 0.1; this agrees well with our simulation data.

\section{Discussion}
\label{sec:discussion}
We have used three-dimensional MHD simulations of cloud-wind interactions to show that gas clouds threaded by tangled magnetic field lines may last substantially longer than hydrodynamic simulations predict.
Our results differ from earlier 3D MHD studies \citep[e.g.,][]{Gregori1999,Gregori2000,Shin2008,Li2013} because we initialized our magnetic fields in a more stable configuration and because we included radiative cooling.

We find that the clumping factor and related observable quantities depend on the magnetic configuration; this has implications for a host of astrophysical observations and suggests that hydrodynamic simulations may be inadequate to predict the densities or luminosities of cold gas clouds interacting with hot flows.
Since the initial state for the magnetic field is typically unknown, our results furthermore suggest that it may be necessary to understand how a gas cloud formed in order to predict its future evolution.

The magnetic field in the medium external to the cloud also influences its observable properties, most importantly through the drag force connecting the cloud to its surroundings (\citealt{Dursi2008} present an analogous equation relevant for weaker magnetic fields).
Equation~\ref{eq:dstop} quantifies how strongly this drag force couples gas clouds to their surroundings.

We now briefly apply our results to (i) ``filaments'' of cold gas in galaxy clusters, (ii) to multiphase galactic winds, and (iii) to the G2 cloud in the Galactic center \citep[e.\,g.][]{Gillessen2012}:
\begin{enumerate}
\item For filaments in galaxy clusters, we estimate the ratio $d_{\text{stop}} / R_{\text{filament}} \sim (\rho_{\text{filament}}/\rho_{\text{ICM}}) \times \beta M^2 / (1+\beta M^2) \sim 10^3$.
Since filaments have a size $R_{\text{filament}} \sim 60$\,pc and a characteristic distance from the cluster center $d \sim 30$\,kpc, We expect the filaments are not strongly coupled to the ambient gas.
They should therefore move at close to the free-fall speed \citep[cf.][]{Hatch2006}.

\item Absorption-line spectroscopy of rapidly star-forming galaxies shows ubiquitous evidence for multi-phase galactic winds \citep[e.\,g.][]{Shapley2006,Martin2006,Weiner2009}.
Moreover, spectroscopy of background galaxies and quasars along sightlines through foreground halos of star-forming galaxies demonstrates that the cool gas extends to large galacto-centric distances, comparable in some cases to the virial radius \citep[e.\,g.][]{Ribaudo2011,Rudie2012,Crighton2013,Werk2014}.
These results have been difficult to understand given the expectation that cool gas accelerated by a hot galactic wind (or by radiation pressure) is rapidly mixed by hydrodynamic instabilities.
Our calculations provide a possible resolution of this puzzle and demonstrate that outflowing magnetized cool gas can become co-moving with a hot galactic wind \textit{prior} to its destruction.
This may be critical to understanding the multi-phase structure of gas in galactic halos.

\item Our estimate of the stopping distance for G2 is uncertain because we don't know the magnetic field strength or gas density along its orbit in the Galactic center.
If we assume an equipartition field with $\beta \sim 1$ we find a stopping distance for G2 comparable to the semi-major axis of its orbit.
Deviations from G2's Keplerian orbit may thus be detectable over an orbital timescale  (e.g., \citealt{Pfuhl2014}).
This suggests the interesting possibility of using G2 to constrain the unknown magnetic field strength and gas density at radii of 100s of AU in the Galactic center.
\end{enumerate}

The results in this letter apply to the specific configuration we adopt for the magnetic field.
To show that our results are not finely-tuned, we have repeated one simulation with a weaker internal field ($\beta_{\text{cloud}} = 10$; see figure~\ref{fig:clumping-factor}), and one simulation in which the field is threaded by a coherent field comparable in strength to the tangled internal  field.
We find similar results in both cases, suggesting that our results apply to a range of initial conditions.

In this letter, we have not addressed the crucial physics setting the size scale of the stable cloudlets shown in figure~\ref{fig:mhd-hydro-comp}.
We explore this in a forthcoming companion paper (O'Leary et al., in prep.).
The companion paper also contains a parameter survey over different cloud density ratios, column densities, wind speeds, magnetic field strengths, and magnetic configurations.
This is important because the relative role of cloud acceleration by the drag force  and cloud destruction by instabilities may depend on these parameters.
In future work, we will also explore the astrophysical applications of our results, particularly to G2 and to galactic winds, in more detail.
We have assumed in this letter that the plasma in the external medium is sufficiently collisional for the MHD equations to apply.
This approximation breaks down in some of the environments to which we apply our results.
The impact of this collisionless physics represents a major uncertainty in our results; however, we leave an exploration of its significance to future work.

\section*{Acknowledgments}
M.M. thanks Christoph Pfrommer for an interesting and helpful conversation about these results.
EQ thanks the participants of the Simons Foundation \textit{Galactic Winds: Beyond Phenomenology} meeting for stimulating discussions.
A.-M.M. is supported by the National Aeronautics and Space Administration through Einstein Postdoctoral Fellowship Award Number PF2-130095 issued by the Chandra X-ray Observatory Center, which is operated by the Smithsonian Astrophysical Observatory for and on behalf of the National Aeronautics Space Administration under contract NAS8-03060.
RO was supported by UC Berkeley's Theoretical Astrophysics Center.
M.M. received support from the Thomas and Alison Schneider Chair in Physics at UC Berkeley.
This work was supported in part by NASA ATP Grant 12- ATP12-0183, by NASA grant NNX10AJ96G, by the David and Lucile Packard Foundation, and by a Simons Investigator Award to EQ from the Simons Foundation.
This work used the Extreme Science and Engineering Discovery Environment (XSEDE), which is supported by National Science Foundation grant number ACI-1053575.
Computing time was provided through XSEDE Allocations TG-AST140039 and TG-AST140047.
We made our figures using the open-source program \textsc{Tioga}.
This research has made use of NASA's Astrophysics Data System.

\bibliographystyle{mn2e}
\bibliography{g2}

\begin{thebibliography}{33}
\expandafter\ifx\csname natexlab\endcsname\relax\def\natexlab#1{#1}\fi

\bibitem[{{Abarca} {et~al.}(2014){Abarca}, {S{\c a}dowski}, \&
  {Sironi}}]{Abarca2014}
{Abarca}, D., {S{\c a}dowski}, A., \& {Sironi}, L. 2014, \mnras

\bibitem[{{Anninos} {et~al.}(2012){Anninos}, {Fragile}, {Wilson}, \&
  {Murray}}]{Anninos2012}
{Anninos}, P., {Fragile}, P.~C., {Wilson}, J., \& {Murray}, S.~D. 2012, \apj,
  759, 132

\bibitem[{{Ballone} {et~al.}(2013){Ballone}, {Schartmann}, {Burkert},
  {Gillessen}, {Genzel}, {Fritz}, {Eisenhauer}, {Pfuhl}, \&
  {Ott}}]{Ballone2013}
{Ballone}, A., {Schartmann}, M., {Burkert}, A., {Gillessen}, S., {Genzel}, R.,
  {Fritz}, T.~K., {Eisenhauer}, F., {Pfuhl}, O., \& {Ott}, T. 2013, \apj, 776,
  13

\bibitem[{{Burkert} {et~al.}(2012){Burkert}, {Schartmann}, {Alig}, {Gillessen},
  {Genzel}, {Fritz}, \& {Eisenhauer}}]{Burkert2012}
{Burkert}, A., {Schartmann}, M., {Alig}, C., {Gillessen}, S., {Genzel}, R.,
  {Fritz}, T.~K., \& {Eisenhauer}, F. 2012, \apj, 750, 58

\bibitem[{{Cooper} {et~al.}(2009){Cooper}, {Bicknell}, {Sutherland}, \&
  {Bland-Hawthorn}}]{Cooper2009}
{Cooper}, J.~L., {Bicknell}, G.~V., {Sutherland}, R.~S., \& {Bland-Hawthorn},
  J. 2009, \apj, 703, 330

\bibitem[{{Crighton} {et~al.}(2013){Crighton}, {Hennawi}, \&
  {Prochaska}}]{Crighton2013}
{Crighton}, N.~H.~M., {Hennawi}, J.~F., \& {Prochaska}, J.~X. 2013, \apjl, 776,
  L18

\bibitem[{{Dursi} \& {Pfrommer}(2008)}]{Dursi2008}
{Dursi}, L.~J., \& {Pfrommer}, C. 2008, \apj, 677, 993

\bibitem[{{Gardiner} \& {Stone}(2008)}]{Gardiner2008}
{Gardiner}, T.~A., \& {Stone}, J.~M. 2008, Journal of Computational Physics,
  227, 4123

\bibitem[{{Gillessen} {et~al.}(2012){Gillessen}, {Genzel}, {Fritz}, {Quataert},
  {Alig}, {Burkert}, {Cuadra}, {Eisenhauer}, {Pfuhl}, {Dodds-Eden}, {Gammie},
  \& {Ott}}]{Gillessen2012}
{Gillessen}, S., {Genzel}, R., {Fritz}, T.~K., {Quataert}, E., {Alig}, C.,
  {Burkert}, A., {Cuadra}, J., {Eisenhauer}, F., {Pfuhl}, O., {Dodds-Eden}, K.,
  {Gammie}, C.~F., \& {Ott}, T. 2012, \nat, 481, 51

\bibitem[{{Gregori} {et~al.}(1999){Gregori}, {Miniati}, {Ryu}, \&
  {Jones}}]{Gregori1999}
{Gregori}, G., {Miniati}, F., {Ryu}, D., \& {Jones}, T.~W. 1999, \apjl, 527,
  L113

\bibitem[{{Gregori} {et~al.}(2000){Gregori}, {Miniati}, {Ryu}, \&
  {Jones}}]{Gregori2000}
---. 2000, \apj, 543, 775

\bibitem[{{Guillochon} {et~al.}(2014){Guillochon}, {Loeb}, {MacLeod}, \&
  {Ramirez-Ruiz}}]{Guillochon2014}
{Guillochon}, J., {Loeb}, A., {MacLeod}, M., \& {Ramirez-Ruiz}, E. 2014, \apjl,
  786, L12

\bibitem[{{Hatch} {et~al.}(2006){Hatch}, {Crawford}, {Johnstone}, \&
  {Fabian}}]{Hatch2006}
{Hatch}, N.~A., {Crawford}, C.~S., {Johnstone}, R.~M., \& {Fabian}, A.~C. 2006,
  \mnras, 367, 433

\bibitem[{{Johansson} \& {Ziegler}(2013)}]{Johansson2013}
{Johansson}, E.~P.~G., \& {Ziegler}, U. 2013, \apj, 766, 45

\bibitem[{{Klein} {et~al.}(1990){Klein}, {Colella}, \& {McKee}}]{Klein1990}
{Klein}, R.~I., {Colella}, P., \& {McKee}, C.~F. 1990, in Astronomical Society
  of the Pacific Conference Series, Vol.~12, The Evolution of the Interstellar
  Medium, ed. L.~{Blitz}, 117--136

\bibitem[{{Klein} {et~al.}(1994){Klein}, {McKee}, \& {Colella}}]{Klein1994}
{Klein}, R.~I., {McKee}, C.~F., \& {Colella}, P. 1994, \apj, 420, 213

\bibitem[{{Li} {et~al.}(2013){Li}, {Frank}, \& {Blackman}}]{Li2013}
{Li}, S., {Frank}, A., \& {Blackman}, E.~G. 2013, \apj, 774, 133

\bibitem[{{Mac Low} {et~al.}(1994){Mac Low}, {McKee}, {Klein}, {Stone}, \&
  {Norman}}]{MacLow1994}
{Mac Low}, M.-M., {McKee}, C.~F., {Klein}, R.~I., {Stone}, J.~M., \& {Norman},
  M.~L. 1994, \apj, 433, 757

\bibitem[{{Martin}(2006)}]{Martin2006}
{Martin}, C.~L. 2006, \apj, 647, 222

\bibitem[{{Miniati} {et~al.}(1999){Miniati}, {Ryu}, {Ferrara}, \&
  {Jones}}]{Miniati1999}
{Miniati}, F., {Ryu}, D., {Ferrara}, A., \& {Jones}, T.~W. 1999, \apj, 510, 726

\bibitem[{{Pfuhl} {et~al.}(2014){Pfuhl}, {Gillessen}, {Eisenhauer}, {Genzel},
  {Plewa}, {Ott}, {Ballone}, {Schartmann}, {Burkert}, {Fritz}, {Sari},
  {Steinberg}, \& {Madigan}}]{Pfuhl2014}
{Pfuhl}, O., {Gillessen}, S., {Eisenhauer}, F., {Genzel}, R., {Plewa}, P.~M.,
  {Ott}, T., {Ballone}, A., {Schartmann}, M., {Burkert}, A., {Fritz}, T.~K.,
  {Sari}, R., {Steinberg}, E., \& {Madigan}, A.-M. 2014, ArXiv e-prints

\bibitem[{{Ribaudo} {et~al.}(2011){Ribaudo}, {Lehner}, {Howk}, {Werk}, {Tripp},
  {Prochaska}, {Meiring}, \& {Tumlinson}}]{Ribaudo2011}
{Ribaudo}, J., {Lehner}, N., {Howk}, J.~C., {Werk}, J.~K., {Tripp}, T.~M.,
  {Prochaska}, J.~X., {Meiring}, J.~D., \& {Tumlinson}, J. 2011, \apj, 743, 207

\bibitem[{{Rudie} {et~al.}(2012){Rudie}, {Steidel}, {Trainor}, {Rakic},
  {Bogosavljevi{\'c}}, {Pettini}, {Reddy}, {Shapley}, {Erb}, \&
  {Law}}]{Rudie2012}
{Rudie}, G.~C., {Steidel}, C.~C., {Trainor}, R.~F., {Rakic}, O.,
  {Bogosavljevi{\'c}}, M., {Pettini}, M., {Reddy}, N., {Shapley}, A.~E., {Erb},
  D.~K., \& {Law}, D.~R. 2012, \apj, 750, 67

\bibitem[{{Ruszkowski} {et~al.}(2007){Ruszkowski}, {En{\ss}lin}, {Br{\"u}ggen},
  {Heinz}, \& {Pfrommer}}]{Ruszkowski2007}
{Ruszkowski}, M., {En{\ss}lin}, T.~A., {Br{\"u}ggen}, M., {Heinz}, S., \&
  {Pfrommer}, C. 2007, \mnras, 378, 662

\bibitem[{{Saitoh} {et~al.}(2014){Saitoh}, {Makino}, {Asaki}, {Baba}, {Komugi},
  {Miyoshi}, {Nagao}, {Takahashi}, {Takeda}, {Tsuboi}, \&
  {Wakamatsu}}]{Saitoh2014}
{Saitoh}, T.~R., {Makino}, J., {Asaki}, Y., {Baba}, J., {Komugi}, S.,
  {Miyoshi}, M., {Nagao}, T., {Takahashi}, M., {Takeda}, T., {Tsuboi}, M., \&
  {Wakamatsu}, K.-i. 2014, \pasj, 66, 1

\bibitem[{{S{\c a}dowski} {et~al.}(2013){S{\c a}dowski}, {Narayan}, {Sironi},
  \& {{\"O}zel}}]{Sadowski2013b}
{S{\c a}dowski}, A., {Narayan}, R., {Sironi}, L., \& {{\"O}zel}, F. 2013,
  \mnras, 433, 2165

\bibitem[{{Schartmann} {et~al.}(2012){Schartmann}, {Burkert}, {Alig},
  {Gillessen}, {Genzel}, {Eisenhauer}, \& {Fritz}}]{Schartmann2012}
{Schartmann}, M., {Burkert}, A., {Alig}, C., {Gillessen}, S., {Genzel}, R.,
  {Eisenhauer}, F., \& {Fritz}, T.~K. 2012, \apj, 755, 155

\bibitem[{{Shapley} {et~al.}(2006){Shapley}, {Steidel}, {Pettini},
  {Adelberger}, \& {Erb}}]{Shapley2006}
{Shapley}, A.~E., {Steidel}, C.~C., {Pettini}, M., {Adelberger}, K.~L., \&
  {Erb}, D.~K. 2006, \apj, 651, 688

\bibitem[{{Shin} {et~al.}(2008){Shin}, {Stone}, \& {Snyder}}]{Shin2008}
{Shin}, M.-S., {Stone}, J.~M., \& {Snyder}, G.~F. 2008, \apj, 680, 336

\bibitem[{{Stone} {et~al.}(2008){Stone}, {Gardiner}, {Teuben}, {Hawley}, \&
  {Simon}}]{Stone2008}
{Stone}, J.~M., {Gardiner}, T.~A., {Teuben}, P., {Hawley}, J.~F., \& {Simon},
  J.~B. 2008, \apjs, 178, 137

\bibitem[{{Stone} \& {Norman}(1992)}]{Stone1992}
{Stone}, J.~M., \& {Norman}, M.~L. 1992, \apjl, 390, L17

\bibitem[{{Weiner} {et~al.}(2009){Weiner}, {Coil}, {Prochaska}, {Newman},
  {Cooper}, {Bundy}, {Conselice}, {Dutton}, {Faber}, {Koo}, {Lotz}, {Rieke}, \&
  {Rubin}}]{Weiner2009}
{Weiner}, B.~J., {Coil}, A.~L., {Prochaska}, J.~X., {Newman}, J.~A., {Cooper},
  M.~C., {Bundy}, K., {Conselice}, C.~J., {Dutton}, A.~A., {Faber}, S.~M.,
  {Koo}, D.~C., {Lotz}, J.~M., {Rieke}, G.~H., \& {Rubin}, K.~H.~R. 2009, \apj,
  692, 187

\bibitem[{{Werk} {et~al.}(2014){Werk}, {Prochaska}, {Tumlinson}, {Peeples},
  {Tripp}, {Fox}, {Lehner}, {Thom}, {O'Meara}, {Ford}, {Bordoloi}, {Katz},
  {Tejos}, {Oppenheimer}, {Dav{\'e}}, \& {Weinberg}}]{Werk2014}
{Werk}, J.~K., {Prochaska}, J.~X., {Tumlinson}, J., {Peeples}, M.~S., {Tripp},
  T.~M., {Fox}, A.~J., {Lehner}, N., {Thom}, C., {O'Meara}, J.~M., {Ford},
  A.~B., {Bordoloi}, R., {Katz}, N., {Tejos}, N., {Oppenheimer}, B.~D.,
  {Dav{\'e}}, R., \& {Weinberg}, D.~H. 2014, ArXiv e-prints

\end{thebibliography}

\label{lastpage}
\end{document}